\documentclass{vgtc}                          

\graphicspath{{figures/}{pictures/}{images/}{./}} 

\usepackage{times}                     

\usepackage{tabu}                      
\usepackage{booktabs}                  
\usepackage{lipsum}                    
\usepackage{mwe}                       
\usepackage{gensymb}
\usepackage{siunitx}

\usepackage{mathptmx}                  

\usepackage[dvipsnames]{xcolor}

\onlineid{0}

\vgtccategory{Research}

\vgtcinsertpkg

\usepackage{cleveref} 

\newcommand{\sysname}{AlloyLens\xspace}

\title{\sysname: A Visual Analytics Tool for High-throughput Alloy Screening and Inverse Design}

\author{
Suyang Li\thanks{e-mail: suyang.li@tufts.edu}\\ %
    \scriptsize Tufts University %
\and Fernando Fajardo-Rojas\\
     \scriptsize Colorado School of Mines %
\and Diego Gomez-Gualdron\\
     \scriptsize Colorado School of Mines %
\and Remco Chang\\
     \scriptsize Tufts University %
\and Mingwei Li\thanks{e-mail: mingwei.li@tufts.edu}\\ %
     \scriptsize Tufts University %
}

\abstract{
     Designing multi-functional alloys requires exploring high-dimensional composition-structure-property spaces, yet current tools are limited to low-dimensional projections and offer limited support for sensitivity or multi-objective tradeoff reasoning. We introduce \sysname, an interactive visual analytics system combining a coordinated scatterplot matrix (SPLOM), dynamic parameter sliders, sensitivity analysis via surrogate model regression curves, and nearest neighbor recommendations. This integrated approach reveals latent structure in simulation data, exposes the local impact of compositional changes, and highlights tradeoffs when exact matches are absent. We validate the system through case studies co-developed with domain experts spanning structural, thermal, and electrical alloy design. 
} 

\begin{document}


\firstsection{Introduction}

\maketitle

Materials discovery and design is inherently exploratory, especially for materials scientists navigating complex design spaces. Beyond automated results, validating and understanding multiple candidate solutions is essential for effective materials design.

\smallskip\noindent We contribute: 1) \sysname, a tightly integrated SPLOM and surrogate model interface; 2) gradient-based local sensitivity analysis with fallback recommendations; and 3) collaborative evaluation with domain experts, demonstrated across real-world use cases.
\subsection{Problem Statement}

Given the desired target properties, the \textbf{inverse design problem} involves identifying the appropriate material compositions or processing parameters to achieve them.
The exploratory nature of the task requires domain experts to iteratively refine hypotheses, compare tradeoffs, and validate candidates across high-dimensional design spaces. 

\subsection{Design Requirement}

Working with our materials science domain experts, we formalize the design requirements of a visualization system to support the inverse design problem as follows:

\smallskip
\noindent{\textbf{(R1) Multi-stage/hierarchical exploration}}
The system must support multi-stage, hierarchical exploration of the design space, allowing users to iteratively and progressively narrow down from broad trends to specific candidate materials.

\smallskip
\noindent{\textbf{(R2) Target range filtering}}
The system must allow users to filter materials based on user-defined target ranges for key properties.

\smallskip
\noindent{\textbf{(R3) Reveal nonlinear relationships}}
The system must visualize nonlinear relationships among alloy composition, intermediate parameters, and resulting material properties.

\smallskip
\noindent{\textbf{(R4) Sensitivity analysis}}
The system must support sensitivity analysis to quantify how changes in composition or processing parameters affect material properties.

\section{Approach}

We implemented our system as an ipywidgets-based Jupyter Notebook widget, allowing users to drive visual analytics directly in their notebook environment and preserve filtered candidate lists from one analysis stage to the next. 
Our interface couples a coordinated SPLOM with parameter sliders, distance-based nearest neighbor highlighting, and real-time gradient-based sensitivity curves to support interactive, multi-dimensional alloy design exploration. 
\subsection{System Architecture}
\textbf{Front-end interface:} 
We built the interface using anywidget~\cite{manz2024anywidget}, ipywidgets~\cite{interactive_Jupyter_widgets}, D3.js~\cite{bostock2011d3}, and regl~\cite{regl} to balance usability, flexibility, and performance. 
Ipywidgets embeds responsive controls in Jupyter for iterative, exploratory workflows (R1); 
D3.js supports customizable, interactive SPLOM views; 
and regl provides GPU-accelerated rendering for real-time, scalable visualizations.


\smallskip\noindent\textbf{Backend computation:} 
To capture nonlinear composition-property relationships and enable real-time sensitivity analysis, we use a lightweight MLP surrogate with PReLU activation, 12 input dimensions, two 1024-unit hidden layers, and 20 outputs (14 properties + 6 features used in the case studies). Trained on all $324,632$ simulated alloys for $\sim 20$ minutes on an Apple Silicon M3 laptop, it achieves an average max normalized residual of $0.6613$. Gradients and predictions from the trained model power interactive sensitivity curves with minimal latency.

\smallskip\noindent\textbf{Data pipeline:} 
To enable responsive alloy exploration, the system loads the 320k-row, 70-attribute dataset into memory, subsampling 20k rows for interactivity. 
Attributes are semantically grouped (scrap inputs, elemental ratios, microstructures, and properties), and fully missing columns --- typically from intermetallic phases that do not form under rapid Scheil solidification --- are zero-filled~\cite{bugelnig2024dataset}. 
Preprocessing is done in Python (pandas) for seamless integration into Jupyter workflows.

\subsection{System Features}
\begin{figure*}
    \centering
    \includegraphics[width=1\linewidth]{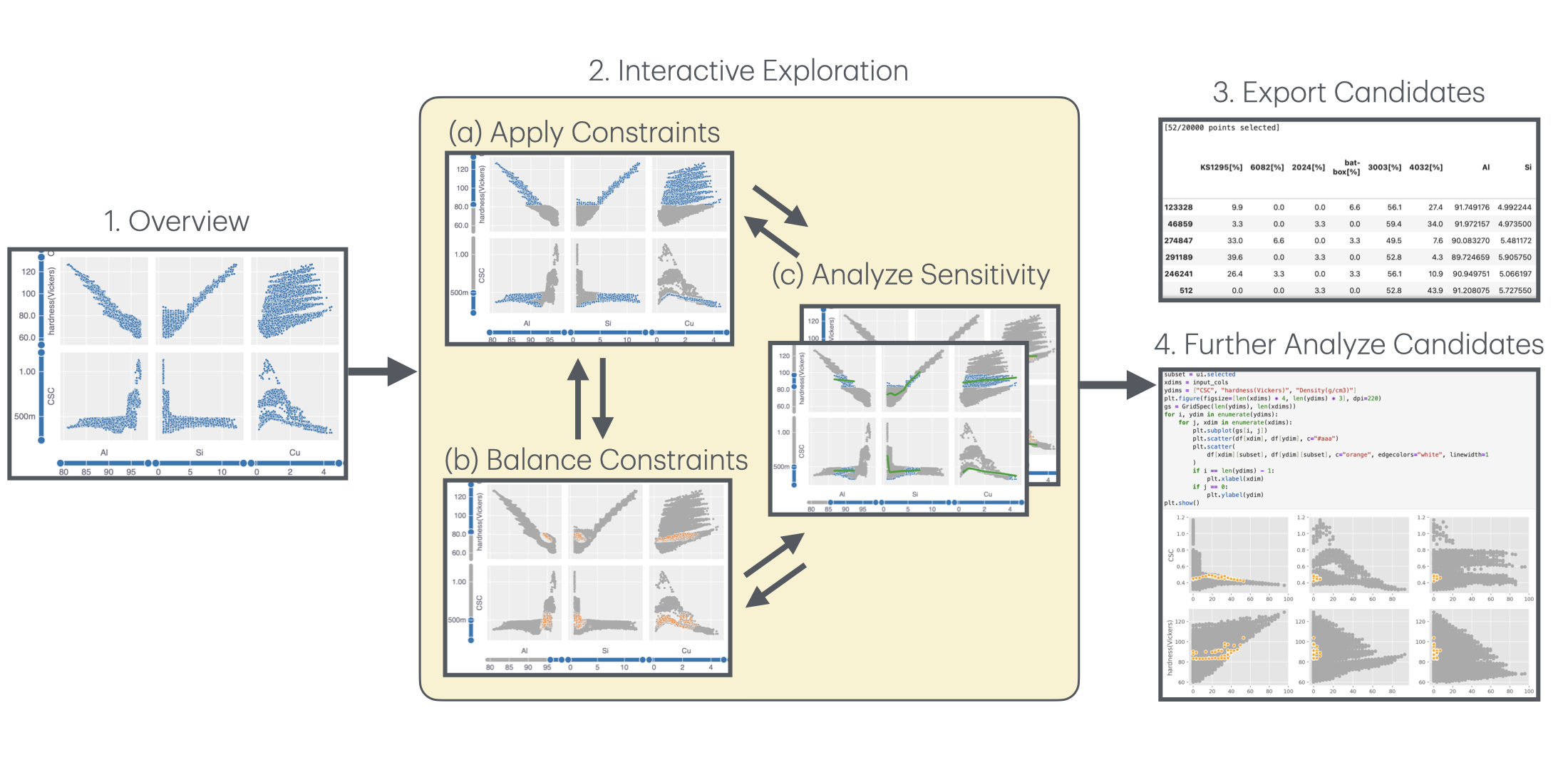}
    \caption{Workflow overview. (1) The user begins with an overview of all candidate materials, shown as chemical compositions mapped to various material properties. (2) They interactively explore the design space by: (2a) applying constraints to filter for desired property ranges; (2b) balancing tradeoffs when constraints conflict; and (2c) analyzing local sensitivity to decide whether further simulation or experimentation is needed.  Finally, users can (3) export selected candidates for (4) further analysis or experimental validation.}
    \label{fig:workflow}
\end{figure*}

\begin{figure*}
    \centering
    \includegraphics[width=1\linewidth]{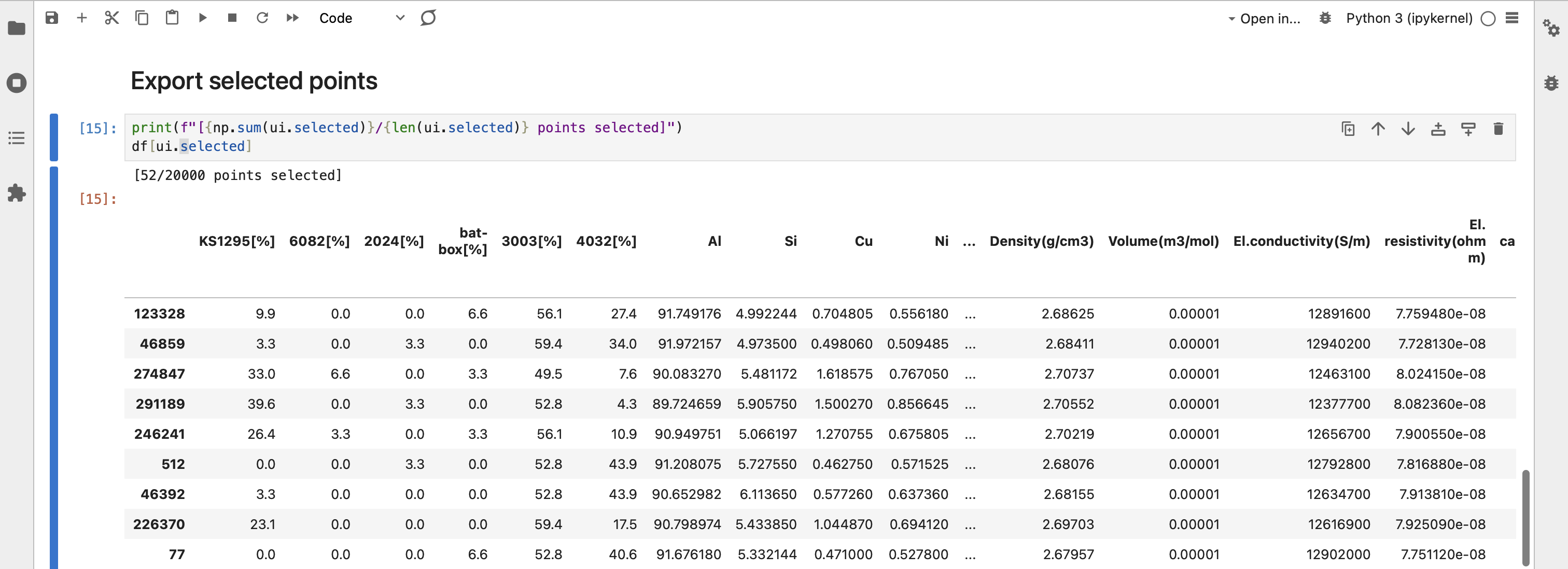}
    \caption{Extract selected data points from the \sysname interface and view them as a pandas DataFrame.}
    \label{fig:export-1}
\end{figure*}

\begin{figure*}
    \centering
    \includegraphics[width=1\linewidth]{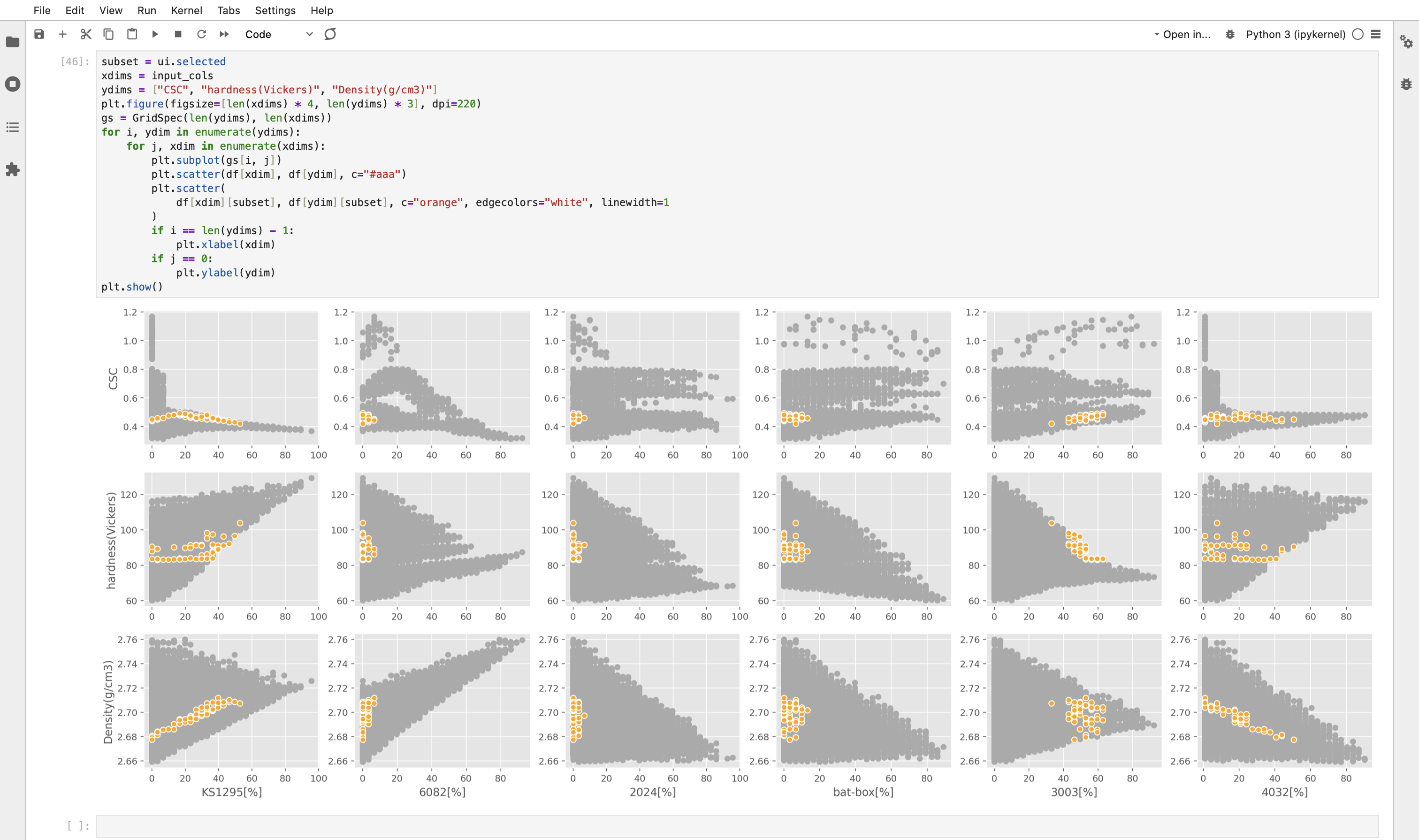}
    \caption{\sysname lets users export selected data points for closer inspection --- for instance, to plot scrap material composition against key alloy properties.}
    \label{fig:export-2}
\end{figure*}

\cref{fig:workflow} presents an overview of the system workflow and interface design. 
Users begin with a global SPLOM view to explore how composition relates to material properties. 
The interface maintains constant visibility of both input controls and output visualizations, reinforcing the causal linkage between compositional adjustments and resultant alloy properties. 
Users iteratively apply property filters, resolve tradeoffs when constraints conflict, and examine property sensitivities via instantaneous slider-SPLOM coupling. This tight feedback loop supports intuitive yet rigorous hypothesis-driven exploration and helps determine whether further simulation or experimentation is needed. Once candidate materials are identified, users can export the selections for downstream computational, visual, or experimental analysis and validation (\cref{fig:export-1,fig:export-2}).

\subsubsection{Overview}
\begin{figure*}
    \centering
    \includegraphics[width=1\linewidth]{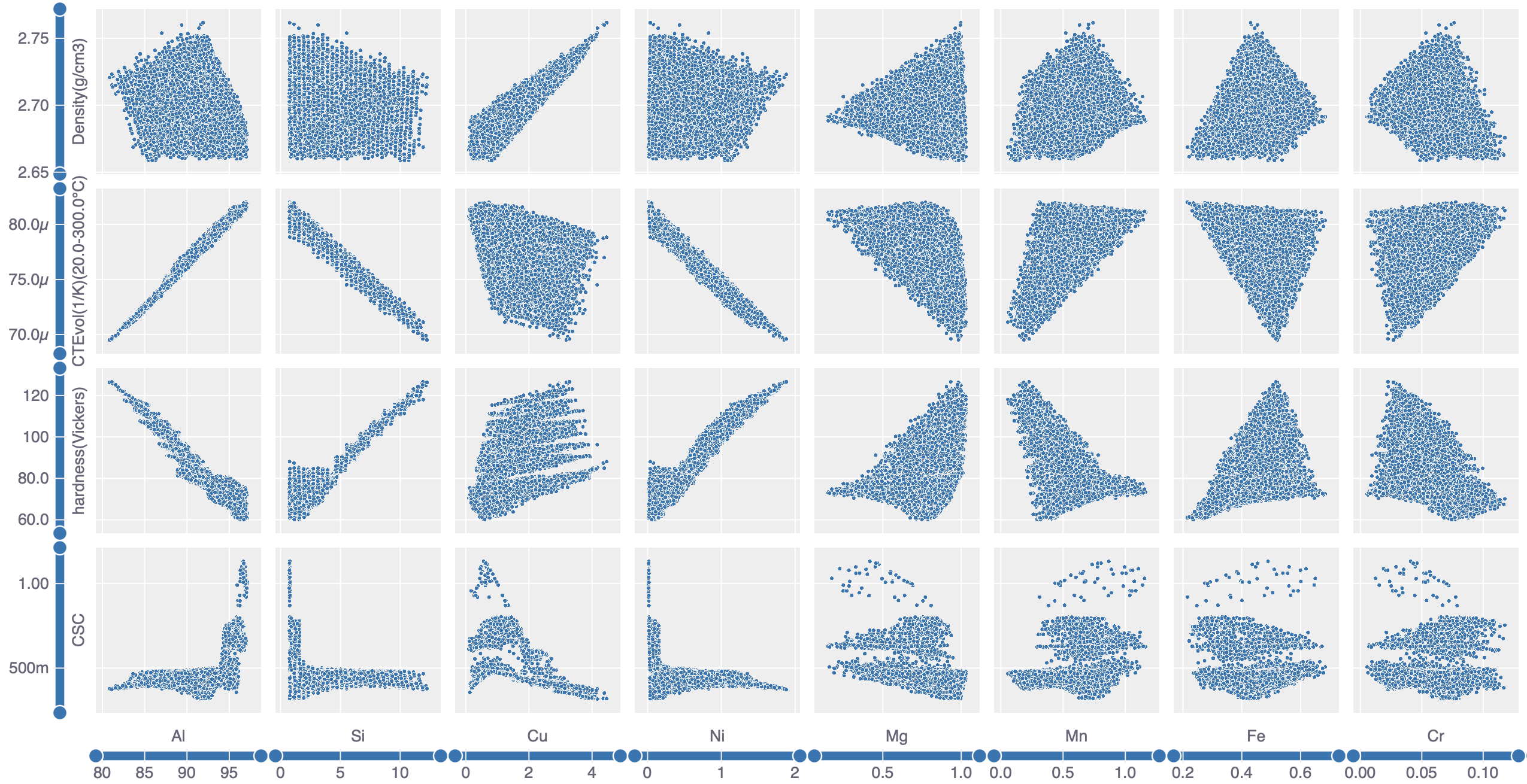}
    \caption{SPLOM overview for hierarchical exploration (R1). The interface starts with a scatterplot matrix showing pairwise relationships between elements and properties, allowing users to identify trends and focus areas for further filtering.}
    \label{fig:overview}
\end{figure*}

To satisfy R1 (hierarchical workflow), the interface presents a SPLOM overview of the composition-property space, where each cell shows 2D projections (e.g., Cu vs. hardness, or Fe vs. density) to reveal correlations and data patterns (\autoref{fig:overview}).
\subsubsection{Feature Selection and Target Sliders}
\begin{figure*}
    \centering
    \includegraphics[width=1\linewidth]{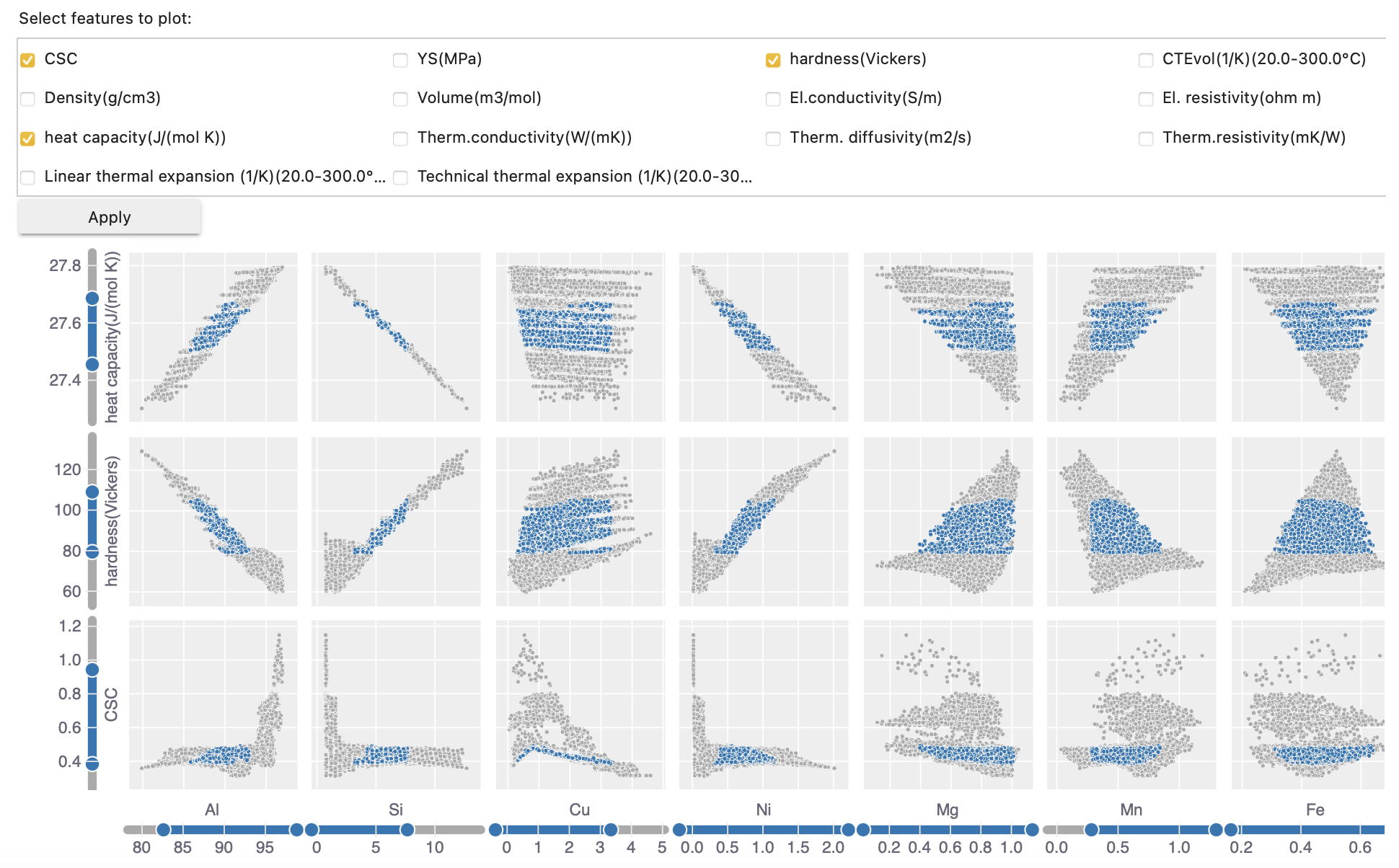}
    \caption{Interactive feature selection. Users define target properties via checkboxes, enabling task-specific filtering and real-time exploration of alloy candidates.}
    \label{fig:checkbox}
\end{figure*}

To satisfy R2 --- configuring user-defined property subsets and target ranges --- \sysname provides an integrated filtering panel. Users can select task-relevant properties (e.g. mechanical properties) directly using the checkboxes above the SPLOM 
and adjust their desired value ranges using interactive sliders aligned with each axis 
(\cref{fig:checkbox,fig:case1-1} top). 
This enables task-specific filtering and real-time refinement of candidate alloys, supporting interpretable exploration tailored to different design goals.

\subsubsection{Nearest Neighbor Recommendation}
When no material satisfies all user-specified property bounds, the system defaults to a distance-based nearest neighbor recommendation. Each property dimension is first normalized to \([0,1]\), and the Euclidean distance from the user-defined target vector is computed for all alloys in the dataset. The top-$k$ nearest candidates --- those with the smallest distances --- are visualized on the SPLOM using a continuous orange color gradient: more saturated hues indicate closer matches to the target, while desaturated tones represent more distant neighbors. This highlights alloys that most closely approximate the user's requirements 
(\autoref{fig:case1-1} bottom).  
\subsubsection{Gradient-Based Sensitivity Analysis}
\begin{figure*}
    \centering
    \includegraphics[width=1\linewidth]{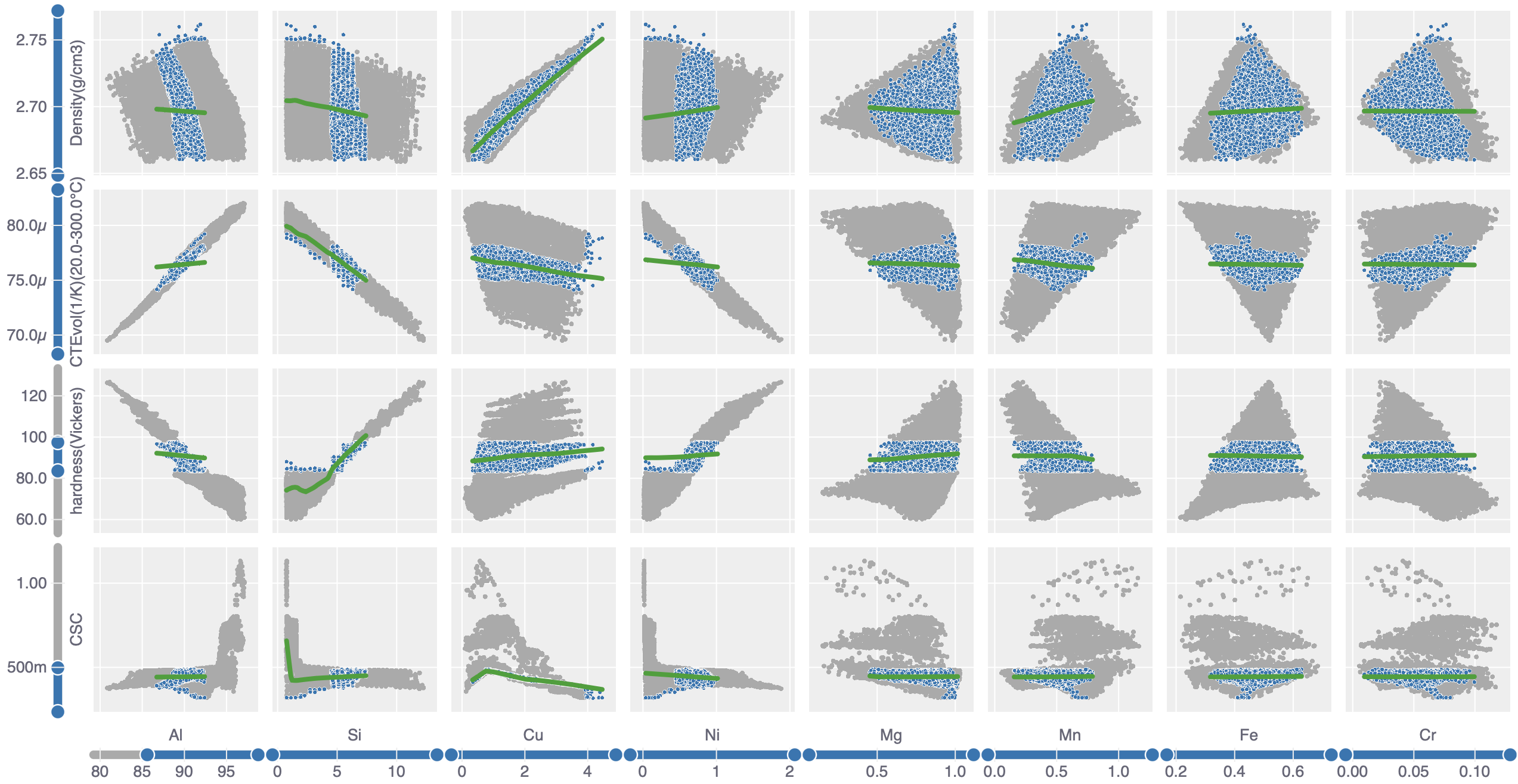}
    \caption{Gradient-based sensitivity analysis (R3, R4). Sensitivity curves derived from an MLP surrogate reveal nonlinear input-output relationships and local sensitivity as users adjust sliders, enabling local tradeoff analysis.}
    \label{fig:filter-curve}
\end{figure*}

To facilitate R3 (reveal nonlinear relationships) and R4 (sensitivity analysis), \sysname visualizes local sensitivity by computing partial derivatives from a trained MLP surrogate.
When a user adjusts an element fraction or property slider, the system starts from the center of the sample compositions, moves along each element axis, queries the MLP for predicted responses, and overlays sensitivity curves on the relevant SPLOM cells (\autoref{fig:filter-curve}).
These curves update instantaneously as the slider moves, thereby revealing local property sensitivities and nonlinearities, enabling precise compositional tuning and tradeoff reasoning. 

\subsubsection{Soft-Match Tolerance Margin} 
\begin{figure*}
    \centering
    \includegraphics[width=1\linewidth]{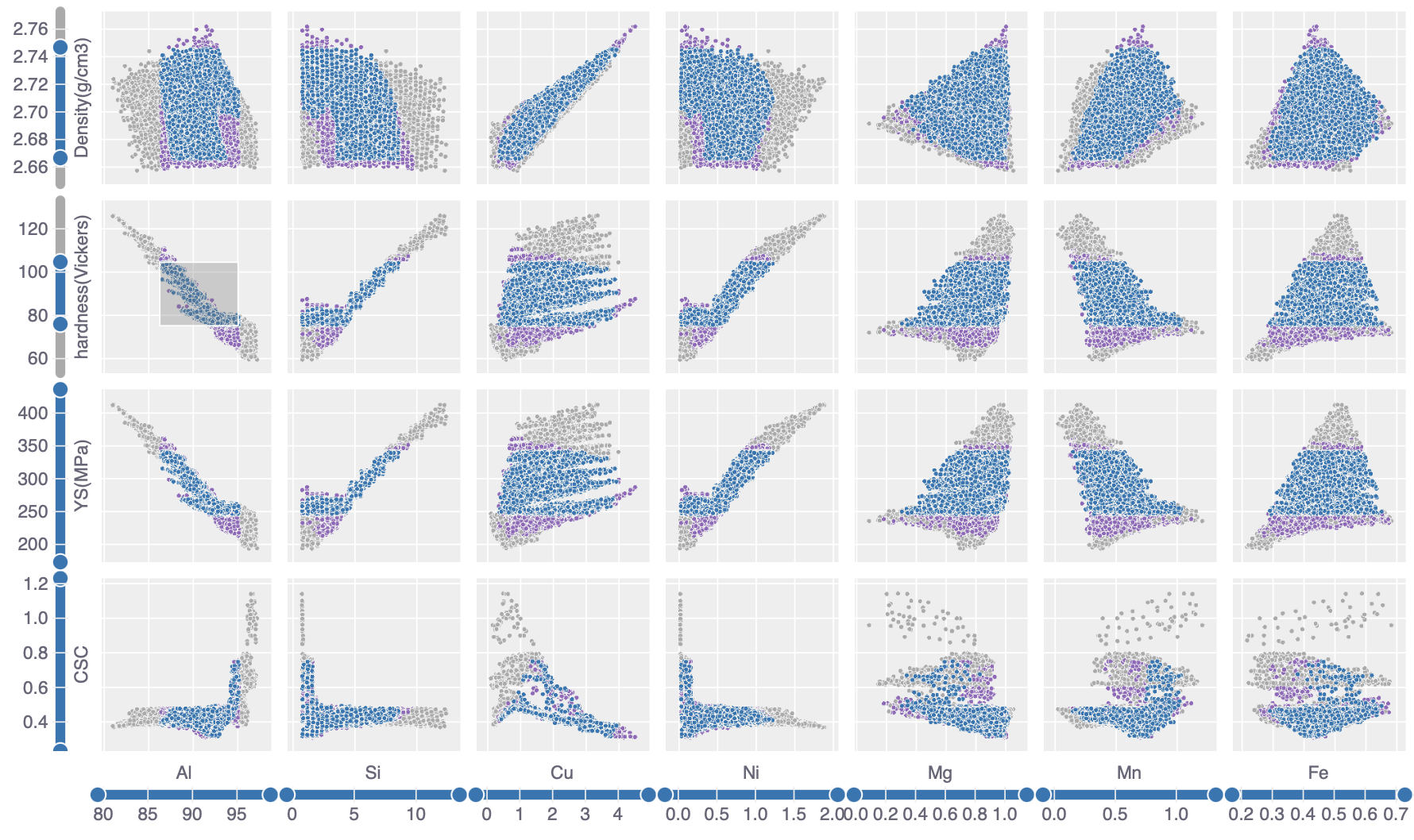}
    \caption{Brush with soft-match tolerance margin. Blue points satisfy all active slider bounds; purple points fall within $\pm 5 \%$ of specified bounds.}
    \label{fig:tolerance}
\end{figure*}

To further support refinement when exact matches are sparse, our system introduces a soft-match tolerance mode. Using brushing interactions, users can visualize near-miss candidates --- those within 5\% of the target bounds --- rendered in purple (\autoref{fig:tolerance}). This feature is particularly useful when working within tight design constraints or low-density regions of the parameter space.

\section{Results}
We demonstrate \sysname with two structural and thermal alloy design tasks selected in collaboration with domain experts. These case studies show how users interactively filter, explore tradeoffs, and identify promising candidates using sensitivity analysis and nearest-neighbor tools.  


\subsection{Structural Alloys for Automotive and Aerospace}

In automotive and aerospace engineering, the selection of materials plays a critical role in ensuring the safety, reliability, and performance of structural components. These industries demand alloys that not only meet stringent mechanical performance standards but also offer weight reduction to enhance fuel efficiency and overall system effectiveness. This case study focuses on the selection of alloys suitable for structural components in automotive and aerospace applications, where high mechanical performance and low weight are paramount to meeting operational and regulatory requirements.
Key property constraints for structural alloys in these applications include: 
\textbf{Yield Strength (YS)} greater than 200 MPa for automotive and 300 MPa for aerospace applications, as it is essential for load-bearing parts; 
\textbf{Hardness (Vickers)} in the range of 80–130 HV to resist wear under cyclic loading; 
\textbf{Density} less than 2.75 g/cm\textsuperscript{3} to reduce part weight and improve fuel efficiency; 
\textbf{Crack Susceptibility Coefficient (CSC)} below 0.5 to reduce the risk of hot cracking; 
\textbf{Solidification Range} (\texttt{delta\_T}) below 100\,\textdegree C, as a narrower range supports better casting; 
\textbf{FCC Phase Fraction} (\texttt{Vf\_FCC\_A1}) above 80\% to maintain strength through a continuous aluminum matrix; and 
\textbf{Fe and Si Content} each below 0.5 wt.\% to minimize the formation of brittle intermetallic phases.

\begin{figure*}
    \centering
    \includegraphics[width=0.9\linewidth]{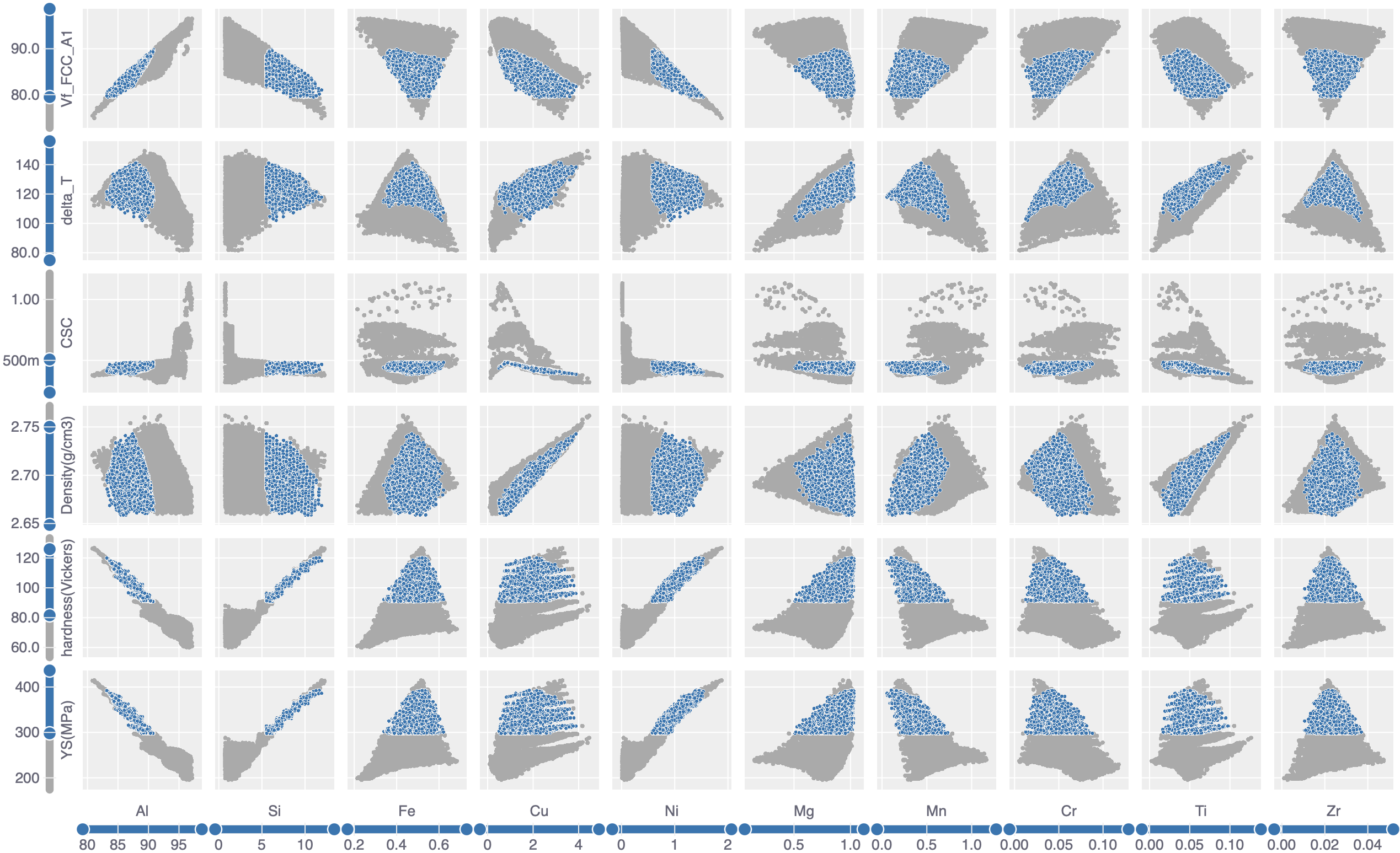}
    
    \vspace{4em}
    \includegraphics[width=0.9\linewidth]{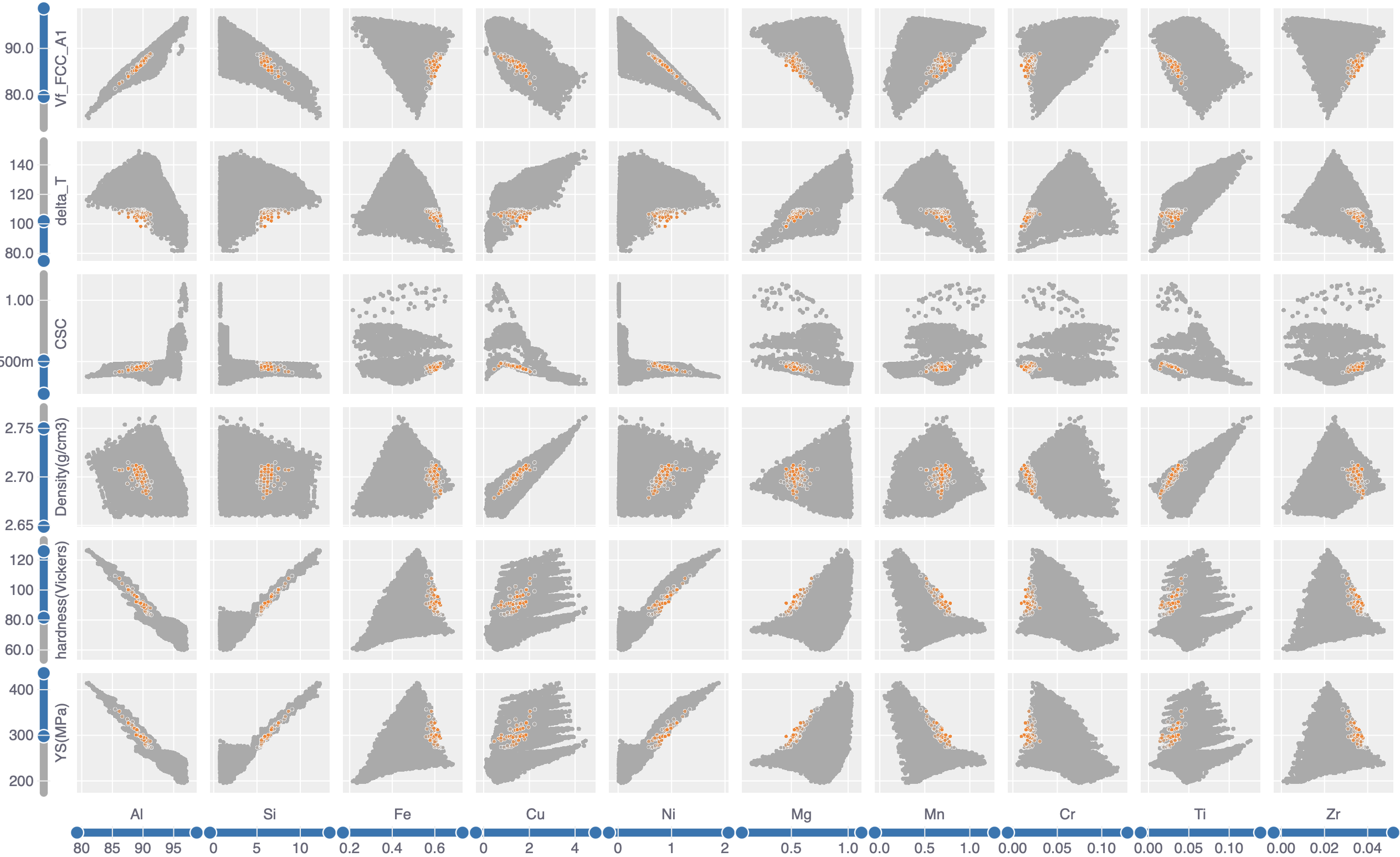}
    \caption{Case 1: Workflow for exploring structural alloys. Users filter for key property targets and inspect candidate clusters,  and use nearest-neighbor suggestions and sensitivity analysis to refine compositions. 
    \textbf{Top}: Before applying the \texttt{delta\_T} constraint, some points meet all other requirements but fall in the upper range of \texttt{delta\_T}, indicating solidification intervals greater than $100\degree C$, which are less favorable for casting.
    \textbf{Bottom}: When the constraint on \texttt{delta\_T} is applied, no points fall within the specified bounds. The system responds by highlighting nearest neighbor solutions using a distinct color.
    Final selections are exported for further analysis.}
    \label{fig:case1-1}
\end{figure*}

\smallskip\noindent\textbf{Workflow} Leveraging \sysname, users first inspect pairwise attribute trends to identify clusters of promising structural alloys. They then apply primary filters for key properties --- yield strength, hardness, density, and CSC according to the aforementioned value bounds (\autoref{fig:case1-1}, top). Next, constraining the solidification range $\Delta T$ below $100 \degree C$, users encounters the scenario where no alloys remain within these tight bounds, automatically activating the nearest neighbor recommendation feature  (\autoref{fig:case1-1}, bottom). 
The user then employs the gradient-based sensitivity curves, interactively adjusting individual element-fraction sliders (e.g. incrementing Si by $+0.1$\ wt.\%) and immediately visualizing the local impact of compositional changes on alloy properties. 
Once satisfied with the refined candidate set, the user exports comprehensive composition-property data of the final selections for downstream analysis and experimental validation (\cref{fig:export-1,fig:export-2}). 
\subsection{Heat Exchanger Materials}

In the design and manufacturing of heat-exchangers, the selection of appropriate materials is crucial to ensure optimal performance while maintaining structural integrity over time. These components are subject to varying thermal loads and must function reliably in challenging environments. As such, heat-exchanger alloys need to strike a balance between thermal, mechanical, and physical properties to meet the demanding requirements.

Heat-exchanger alloys must combine high thermal conductivity ($> 150$ $W/m\cdot k$) with low mass density ($<2.7 \ g/cm^3$), a linear thermal expansion coefficient matched to that of adjoining components ($20$-$26 \times 10^{-6}\ 1/K$) to prevent stress and potential damage during thermal cycling. Moderate hardness ($60$-$100 \ HV$) is required for mounting and supporting the components without causing excessive wear or deformation. gra$1$-$12$ wt.\% Si is needed to ensure thermal conductivity as well as fluidity during casting processes. 

\begin{figure*}
    \centering
    \includegraphics[width=\linewidth]{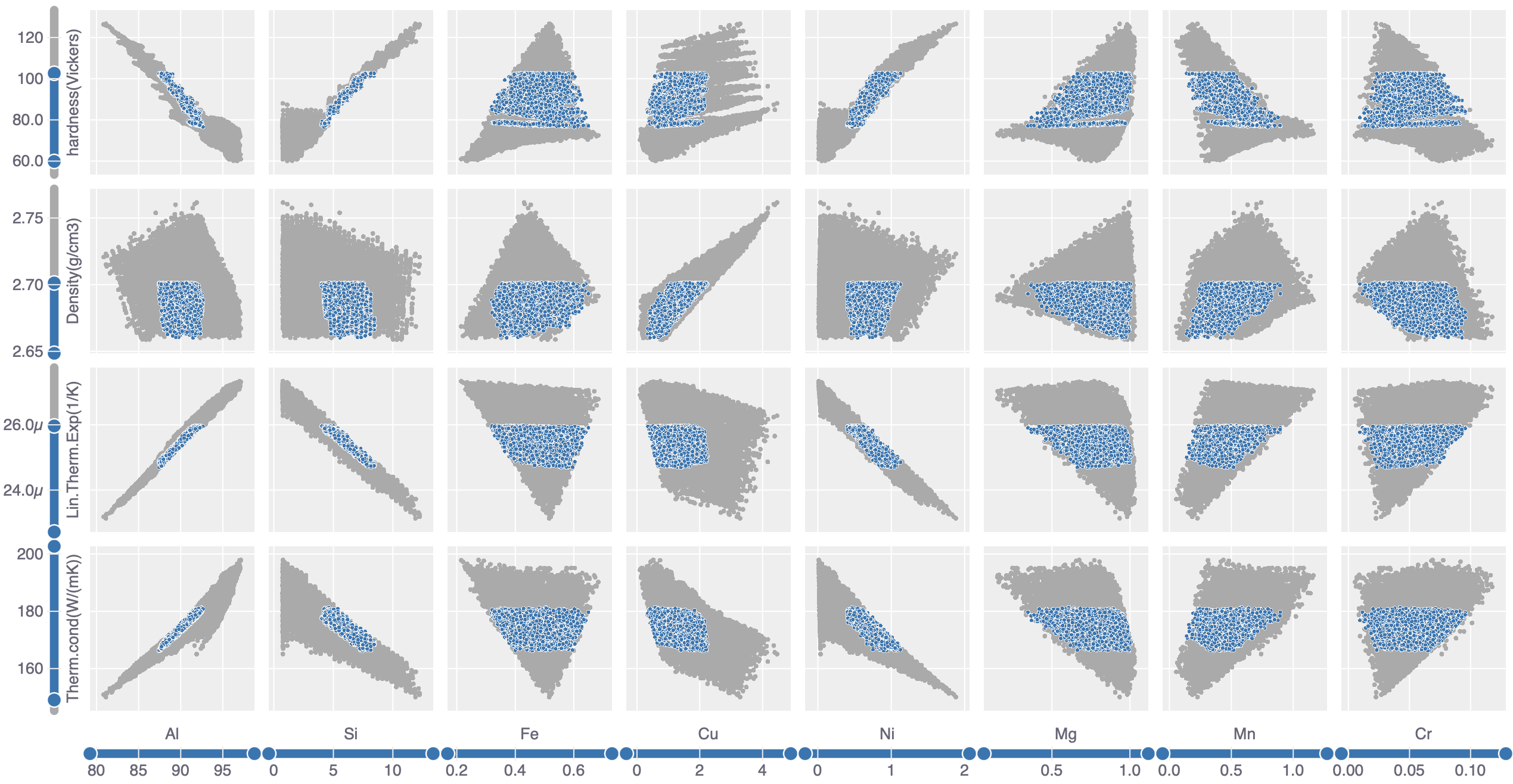}
    \caption{Case 2: Heat Exchanger Materials. The visualization system supports the identification of viable heat-exchanger alloys by allowing users to interactively apply multi-property constraints—including thermal conductivity, density, thermal expansion, hardness, and Si content. As constraints are adjusted, candidates satisfying all requirements are highlighted in real time, enabling efficient exploration and selection. Selected materials can be exported for further analysis or testing.}
    \label{fig:case2}
\end{figure*}

\smallskip\noindent\textbf{Workflow} In this scenario, viable materials exist in the design space, so our tool's main function is to help users seamlessly navigate and interact with the multi-dimensional space and locate the solutions. The user sets the thermal conductivity, density, linear thermal expansion coefficient, and hardness; as sliders are adjusted, all alloy meeting the constraints are highlighted in blue in real time, revealing the viable heat-exchanger candidates (\cref{fig:case2}). All data points can be exported in one click for downstream simulation or experimental testing.

\subsection{Evaluation with Domain Experts}
AlloyLens was developed in close collaboration with materials scientists, who emphasized the desirable properties often conflict and ideal ranges are rarely compatible. They describe the ability to explore variable-range combinations --- not just hard filters --- as critical to identifying feasible solutions. \sysname was valued for its ability to reveal acceptable tradeoff regions and offer candidates that fall close to target constraints. Experts also noted that the surrogate model enabled exploration beyond observed data, describing this as essential for hypothesis refinement when data coverage is limited. Altogether, domain experts noted that  \sysname supported the judgment-based, iterative reasoning that characterizes real-world material design workflows.

\section{Project Links}
\begin{itemize}
    \setlength{\itemsep}{0pt}
    \item PyPI package: https://pypi.org/project/alloylens/
    \item Web demo: http://susiesyli.com/alloylens-web/
    \item GitHub: alloylens
    \item SciVis Contest 2025: https://sciviscontest2025.github.io/data/
\end{itemize}
\acknowledgments{
This work was partially supported by NSF under grant number NSF OAC-2118201.
}

\bibliographystyle{abbrv-doi}

\bibliography{template}



\begin{table*}[ht]
\centering
\small
\begin{tabular}{l r r p{1.5cm} p{1.6cm}}
\toprule
\textbf{Property} & \textbf{Mean} & \textbf{Std Dev} & \textbf{Max Residual (normalized)} & \textbf{Max Residual (original scale)} \\
\midrule
CSC & 0.4562 & 0.0637 & 2.9248 & 0.1863 \\
YS (MPa) & 277.798 & 41.7141 & 0.4356 & 18.1930 \\
Hardness (Vickers) & 84.986 & 12.7542 & 0.4202 & 5.3612 \\
CTEvol (1/K) (20–300$\degree C$) & 7.73e-5 & 2.24e-6 & 0.1727 & 3.87e-7 \\
Density ($g/cm^3$) & 2.6964 & 0.0162 & 0.3264 & 0.0053 \\
Volume ($m^3/mol$) & 1.02e-5 & 3.59e-8 & 0.3051 & 1.10e-8 \\
El. conductivity (S/m) & 1.28e7 & 6.79e5 & 0.3472 & 2.36e5 \\
El. resistivity ($\Omega \cdot m$) & 7.83e-8 & 4.20e-9 & 0.3624 & 1.52e-9 \\
Heat capacity ($J/mol \cdot K$) & 27.6340 & 0.0910 & 0.3168 & 0.0288 \\
Therm. conductivity (W/m·K) & 176.168 & 8.0090 & 0.4686 & 3.7516 \\
Therm. diffusivity ($m^2/s$) & 6.51e-5 & 2.65e-6 & 0.5760 & 1.53e-6 \\
Therm. resistivity (mK/W) & 5.69e-3 & 2.61e-4 & 0.4572 & 1.19e-4 \\
Lin. thermal exp. (1/K) (20–300$\degree C$) & 2.58e-5 & 7.48e-7 & 0.1784 & 1.33e-7 \\
Tech. thermal exp. (1/K) (20–300$\degree C$) & 2.35e-5 & 6.82e-7 & 0.1903 & 1.30e-7 \\
Vf\_FCC\_A1 & 88.4799 & 3.5077 & 1.6533 & 5.8005 \\
$\Delta T$ & 120.2042 & 9.5306 & 0.5101 & 4.8643 \\
T(liq) & 658.3913 & 6.4341 & 0.5376 & 3.4586 \\
Eut. frac. [\%] & 57.3786 & 18.1783 & 1.7062 & 31.0615 \\
Vf\_DIAMOND\_A4 & 3.1701 & 1.9947 & 0.3136 & 0.6254 \\
Vf\_AL15SI2M4 & 2.2810 & 0.5720 & 1.4391 & 0.8235 \\
\bottomrule
\end{tabular}
\caption{Dataset property statistics (mean and standard deviation) alongside surrogate model performance, shown as maximum residuals both normalized and in the original data scale. All surrogate model output properties were normalized to have a standard deviation of 1, ensuring fair comparison and meaningful scaling across outputs.}
\label{table:performance}
\end{table*}

\end{document}